\documentclass[prd,twocolumn,showpacs,preprintnumbers,footinbib,floatfix]{revtex4}

\usepackage{graphicx}
\usepackage[raggedright,hang]{subfigure}
\usepackage{epstopdf}
\usepackage{rotating}
\usepackage{color}
\usepackage{hyperref}
\usepackage{amssymb}
\usepackage{amsmath}
\usepackage{amsfonts}
\usepackage{fancyhdr}
\usepackage{datetime}
\usepackage{times}

\newcommand{\hrs}{\textrm{h}}

\newcommand{\rmd}{\mathrm{d}}
\newcommand{\vecb}[1]{\boldsymbol{\mathrm{#1}}}
 
\newcommand{\eref}[1]{(\ref{#1})}

\newcommand{\Eref}[1]{Eq.~(\ref{#1})}
\newcommand{\Esref}[1]{Eqs.~(\ref{#1})}
\newcommand{\Sref}[1]{Sec.~\ref{#1}}
\newcommand{\Fref}[1]{Fig.~\ref{#1}}

\newcommand{\Hz}{\mathrm{Hz}}

\def\erfc{\mathrm{erfc}}

\newcommand{\fkdot}[1]{f^{(#1)}}

\newcommand{\tnudot}[1]{\dot{\tilde{\nu}}}

\def\Doppler{{\mathrm{p}}}
\def\vDoppler{\vecb\Doppler}

\def\trapo{\mathrm{\tiny T}}
\newcommand{\F}{\mathcal{F}}
\newcommand{\Amp}{\mathcal{A}}
\newcommand{\Astat}{\mathcal{Z}}

\newcommand{\xx}{\textit{\textrm{x}}}

\newcommand{\Tdata}{T_{\tiny\textrm{data}}}
\newcommand{\coho}{\eta}
\newcommand{\statfac}{\mathcal{E}}

\newtimeformat{dottime}{\twodigit{\THEHOUR}:\twodigit{\THEMINUTE}}
\settimeformat{dottime}

\DeclareGraphicsExtensions{.png}

 \def\dcc{LIGO-P1000130-v3}
 \def\aei{AEI-2010-181}

\begin{document}

\pagestyle{fancy}

\preprint{\dcc}
\preprint{\aei}

\rhead[]{}
\lhead[]{}

\title{
Sliding coherence window technique
for hierarchical detection\\ of continuous gravitational waves
}

\author{Holger J. Pletsch}
\email{Holger.Pletsch@aei.mpg.de} 
\affiliation{Max-Planck-Institut f\"ur Gravitationsphysik (Albert-Einstein-Institut)
    and Leibniz Universit\"at Hannover, Callinstra{\ss}e 38, D-30167 Hannover, Germany}

\date{\currenttime, \today}

\begin{abstract}
\noindent
A novel hierarchical search technique is presented for all-sky surveys 
for continuous gravitational-wave sources, such as rapidly spinning nonaxisymmetric 
neutron stars. Analyzing yearlong detector data sets over realistic ranges of parameter space
using fully coherent matched-filtering is computationally prohibitive.
Thus more efficient, so-called hierarchical techniques are essential.
Traditionally, the standard hierarchical approach consists of dividing the data 
into nonoverlapping segments of which each is coherently analyzed
and subsequently the matched-filter outputs from all segments are combined incoherently. 
The present work proposes to break the data into subsegments shorter 
than the desired maximum coherence time span (size of the coherence window). 
Then matched-filter outputs from the different subsegments are efficiently combined 
by sliding the coherence window in time: Subsegments whose timestamps
are closer than coherence window size are combined coherently, 
otherwise incoherently. Compared to the standard scheme at the same
coherence time baseline, data sets longer by about \mbox{$50-100\,\%$} would 
have to be analyzed to achieve the same search sensitivity as with the sliding 
coherence window approach. Numerical simulations attest to the analytically 
estimated improvement.
\end{abstract}


\maketitle

\section{Introduction\label{sec:Introduction}}

Rapidly rotating neutron stars are anticipated to emit 
continuous gravitational-wave (CW) signals 
through various plausible 
scenarios~\cite{owen-1998-58,ushomirsky:2000,cutler:2002-66,jones-2002-331,owen-2005-95,horowitz-kadau-2009}
due to asymmetries. 
While the majority of such neutron stars eludes electromagnetic
observations, their population might potentially  be probed only by means of 
gravitational waves~\cite{NeutronStars2009}. Currently, an international network 
of laser-interferometric \mbox{detectors~\cite{ligoS5,virgo2006,geo2008,tama2004}}
is in operation.
The observational upper limits obtained from known radio 
pulsars~\cite{crab-2008,knownPulS5-2010} and all-sky 
surveys~\cite{pshS4:2008,powerflux:2009,S4EAH,S5R1EAH,eahurl}
already constrain the physics of neutron stars,
and thus a detection of a CW signal from a spinning neutron star would shed 
light on their currently rather uncertain physics~\cite{NSpromises2010}.

Extremely sensitive data analysis techniques are needed to detect prior unknown 
CW sources because of their expected low signal-to-noise ratios.  
A powerful method has been derived~\cite{jks1} 
based on the principle of maximum likelihood detection,
leading to coherent matched filtering. 
CW signals are quasimonochromatic with a slowly changing intrinsic frequency.  
However, the Earth's motion relative to the solar system barycenter (SSB) 
generates a Doppler modulation in amplitude and phase of the waveform 
at a terrestrial detector.  
As shown in~\cite{jks1}, the coherent matched-filtering statistic can be
analytically maximized over the \emph{amplitude parameters} describing 
the signal's amplitude variation.
The so-obtained coherent detection statistic is referred to as
the \mbox{$\F$-statistic}, which can also include
multiple detector data streams~\cite{CutlerMultiIFO}.
Thus, an explicit search (evaluating $\F$) is only done over
the \emph{phase parameters} describing the signal's phase evolution:
the source's sky location, frequency, and frequency derivatives (``spindowns''). 

However, what ultimately limits the search sensitivity in scanning the entire sky
for previously unknown CW sources is the finite computational resources.
For yearlong data sets, searching a realistic portion of parameter space 
is computationally absolutely impractical~\cite{bccs1:1998,jks1}.
This is due to the apparently enormous number of template waveforms needed 
to discretely cover the search parameter space,
increasing as a high power of the coherent integration time.
In consequence, viable all-sky fully coherent \mbox{$\F$-statistic} 
searches are restricted to much shorter coherent integration times,
despite that the $\F$-statistic can be very efficiently computed using the fast 
Fourier transform (FFT) algorithm \cite{jks1,Resamp2010}.

All-sky surveys sifting through yearlong data sets for previously unknown isolated 
CW sources are accomplished by incoherently combining either 
excess power or \mbox{$\F$-statistic} values from shorter segments of data.
In the power-combining methods~\cite{pshS4:2008,powerflux:2009}, the segment 
duration is chosen short enough (typically, $30$ minutes) so that the CW signal 
power resides in a single frequency bin during each segment.
In contrast, so-called ``hierarchical''  \mbox{$\F$-statistic}-based 
methods~\cite{cutler:2005,hough:2005,pletsch:GCT} coherently track the CW 
signal phase over longer segments (typically of the order of a day or a few days).

In this work, the ``standard'' hierarchical detection scheme refers to the following 
approach. Divide the data into nonoverlapping segments of duration~$T$.
Then, for a given point in search parameter space, the 
\mbox{$\F$-statistic} is computed separately for each segment and subsequently
\mbox{$\F$ values} from all segments are summed.
Thus $T$ defines the maximum time span of maintained phase 
coherence to the signal.
This approach is efficient, because computing the coherent matched-filtering 
statistic~$\F$ just over~$T$ allows one to use a \emph{coarse} grid of templates 
in phase parameter space, compared to one required for the entire data set.
Only when incoherently combining the \mbox{$\F$-statistic} results from all 
segments is a common \emph{fine} grid of templates necessary.

Recent progress in understanding the global
correlations~\cite{pletsch:2008,prixitoh:2005} of 
the \mbox{$\F$-statistic} in the phase parameters
has lead to a substantially more sensitive 
hierarchical search technique~\cite{pletsch:GCT}.
This method has addressed a  long-standing problem, 
namely, the design of, and link between, the coarse and fine grids. 
A very useful geometric tool in this context is the concept
of a metric, as first investigated in \cite{Sathy1:1996,owen:1996me},
measuring the fractional loss in expected $\F$-statistic
for a given signal  at a nearby grid point.
While such a metric has been well studied for the coherent stage
\cite{bccs1:1998,ptolemetric,prix:2007mu},
in~\cite{pletsch:GCT} the first analytical metric
for the incoherent combination step has been found
by exploiting new coordinates on the phase parameter space.
This analytical ``semicoherent metric'' has lately been further studied 
and extended to greater generality in~\cite{pletsch:scmetric}.
This technology is currently also implemented and
employed by \mbox{Einstein@Home}~\cite{eahurl}, 
a volunteer distributed computing project carrying 
out the most sensitive all-sky CW surveys.

The present work presents a novel hierarchical 
search strategy 
which builds on the results of~\cite{pletsch:GCT,pletsch:scmetric},
while further enhancing the search sensitivity.
Previous hierarchical search methods divide the data 
into nonoverlapping segments whose length is equal to coherent time baseline~$T$.
Here, a partitioning of the data into segments 
\emph{shorter} than the desired maximum 
coherence length~$T$ is considered and subsequently
one ``slides'' a coherence window of size~$T$ over the segments.
As a result, segments which are closer than coherence window size~$T$
are coherently combined and otherwise incoherently.
This scheme also ensures that the same semicoherent
metric as derived in~\cite{pletsch:GCT,pletsch:scmetric}
governs the  template grid construction in phase parameter space
but considerably enhances the overall search sensitivity.

Section~\ref{sec:CWsignals} briefly recaps the
continuous gravitational-wave signal waveform.
Section~\ref{sec:SemiCohDetSeg} describes
the coherent matched-filtering statistic~$\F$ and 
what we refer to as the ``standard hierarchical search scheme.''
The idea behind the sliding coherence window approach
is elucidated in \Sref{sec:SlideWin}, along
with an analytical sensitivity estimation.
The improved performance is demonstrated
in \Sref{sec:PerfDemo} by means of Monte Carlo simulations.
In addition, \Sref{sec:ComputingCost} compares the estimated sensitivity
at fixed computational cost. Finally,
concluding remarks follow in \Sref{sec:Conclusion}.

\section{Continuous gravitational-wave signals\label{sec:CWsignals}}

The dimensionless signal response function~$h(t)$ of an 
interferometric detector to a weak plane gravitational wave in the 
long-wavelength approximation is a linear combination of the form~\cite{jks1},
\begin{equation}
  h(t) = F_+ (t) \, h_+(t) + F_\times (t) \, h_\times(t) \,.
  \label{e:h_t}
\end{equation}
The antenna pattern functions $F_{+}(t)$ and $F_\times(t)$ are given by
\begin{subequations}
  \label{e:Fpc_t}
\begin{align}
  F_+ (t) &= a(t) \cos 2\psi + b(t) \sin 2\psi \,,\\
  F_\times (t) &= b(t) \cos 2\psi - a(t) \sin 2\psi \,,
\end{align}
\end{subequations}
where $\psi$ represents the polarization angle of the signal, and the angle
between the detector arms is assumed to be $\pi/2$. For explicit
expressions of the functions $a(t)$ and $b(t)$, the reader is
referred to Ref.~\cite{jks1}.

In the case of an isolated, rapidly rotating neutron
star with a nonaxisymmetric deformation and negligible proper motion~(cf. \cite{jk2,jk3}), 
the waveforms corresponding to the plus~($+$) and cross~($\times$) 
polarizations are 
\begin{equation}
  h_+ (t) = A_+ \, \sin \Psi(t)\,, \qquad 
  h_\times (t) = A_\times \, \cos \Psi(t) \,,
  \label{e:h_pc_t}
\end{equation}
where $A_+$ and $A_\times$ are the constant plus and cross polarization
amplitude parameters, respectively, and $\Psi(t)$ is the phase
of the signal. The parameters  $A_+$ and $A_\times$ can be expressed
in terms of the gravitational-wave strain tensor amplitude~$h_0$
and the inclination angle~$\iota$ as
\begin{equation}
 A_+ = h_0\left( 1+\cos^2\iota\right)/2 \,,\qquad
 A_\times = h_0 \cos \iota \, .
\end{equation}
The phase $\Psi(t)$ of the CW signal at detector time~$t$ takes the following form~\cite{jks1}:
\begin{align}
   \Psi(t) &= \Phi_0 + \Phi(t) \nonumber\\
   &= \Phi_0 + 2\pi  \sum_{k=0}^{s} \, \frac{\fkdot{k}(t_0)}{(k+1)!} 
  \,  \left[t-t_0 + \frac{\vec r (t) \cdot \vec n}{c}\right]^{k+1},
  \label{e:Phase1}
\end{align}
where $\Phi_0$ is the initial phase, $\fkdot{0} \equiv f$ denotes the 
frequency, and $\fkdot{k>0}$ is the $k$th frequency time derivative (also called ``spindown''), 
evaluated at the SSB at reference time $t_0$.  
The integer~\mbox{$s>0$} denotes the number of frequency time derivatives to be taken into account;
therefore, it holds $\fkdot{k>s}=0$. The vector $\vec r (t)$ connects from the SSB to the detector, 
 $c$ is the speed of light, and  $\vec n$ is a constant unit vector pointing
from the SSB to the location of the CW source.
The source's sky location is determined by two independent coordinates, for example, one can
use equatorial coordinates of right ascension and
declination, denoted by~$\alpha$ and~$\delta$, respectively. In these coordinates:  
$\vec n = (\cos \delta \, \cos \alpha, \cos \delta \, \sin \alpha, \sin \delta)$. 
The collection of phase parameters will be summarized by
the vector \mbox{$\vDoppler\equiv\left(f,\fkdot{1},...,\fkdot{s}, \alpha,\delta\right)$}.

Using \Esref{e:Fpc_t}, \eref{e:h_pc_t} and \eref{e:Phase1}, it is possible
to rewrite \Eref{e:h_t} as follows:
\begin{equation}
  h(t) = \sum_{\mu=1}^4 \Amp_\mu h_\mu(t) \,,
  \label{e:h_t_2}
\end{equation}
where the four amplitude parameters $(A_+,A_\times,\psi,\Phi_0)$
have been reparametrized by the 4-vector $\Amp \equiv (\Amp_1,\Amp_2,\Amp_3,\Amp_4)$,
whose individual components are
\begin{align}
\Amp_1 &= A_{+}\,\cos2\psi \, \cos\Phi_0 - A_{\times}\sin2\psi\, \sin\Phi_0 \,,
\nonumber\\[1ex]
\Amp_2 &= A_{+}\sin2\psi\, \cos\Phi_0 + A_{\times}\cos2\psi\, \sin\Phi_0\,,
\nonumber\\[1ex]
\Amp_3 &= - A_{+}\cos2\psi\, \sin\Phi_0 - A_{\times}\sin2\psi\, \cos\Phi_0\,,
\nonumber\\[1ex]
\Amp_4 &= - A_{+}\sin2\psi\, \sin\Phi_0 + A_{\times}\cos2\psi\, \cos\Phi_0\,,
\label{e:A1234}
\end{align}
and the functions $h_\mu(t)$ have been defined as
\begin{align}
\begin{array}{c}
h_1(t) = a(t)\cos\Phi(t)\,, \hspace{3mm} h_2(t) = b(t)\cos\Phi(t) \,,
\\[2ex]
h_3(t) = a(t)\sin\Phi(t)\,, \hspace{3mm} h_4(t) = b(t)\sin\Phi(t) \,.
\end{array}
\end{align}

\section{Standard hierarchical detection scheme\label{sec:SemiCohDetSeg}}

At a given detector time~$t$, the detector output data time series is denoted by $x(t)$. 
In the absence of any signal, the data contain only noise~$n(t)$, 
which is assumed to be a zero-mean, stationary, and 
Gaussian random process~\footnote{
Over the typical long coherent integration times of hierarchical CW searches
as considered here, the Gaussian noise assumption can also be well justified 
in practice based on the central limit theorem for the vast majority of frequency 
bands of the real detector output~(cf. the LIGO searches in~\cite{S4EAH,S5R1EAH}).
There is also evidence~\cite{Finn2001} that
the performance of the matched-filtering method for Gaussian noise is also 
satisfactory for the case of non-Gaussian noise.
}. 
When a signal~$h(t)$ is present, the noise is assumed to be additive,
so that \mbox{$x(t) = n(t) + h(t)$}. 

For simplicity, in this work only a single-detector input data stream is considered. 
However, based on the results of~\cite{CutlerMultiIFO} (and also \cite{prix:2007mu}), 
it is straightforward to generalize the proposed search technique 
to multiple-detector input data, as well as to time-varying noise.

To contrast with the sliding coherence window approach, 
this section describes the standard hierarchical detection scheme
which sums one \mbox{$\F$-statistic} value from each of
$N$ \emph{nonoverlapping} segments of duration~$T$. 
For simplicity, in this presentation the data set is taken to be contiguous,
so that the total data time span is written as {$T_{\rm\tiny data}=NT$}.
The individual segments are labeled by the index~\mbox{$j=1,...,N$}.
Let $t_j$ denote the time midpoint of segment~$j$, 
which thus spans the time interval~\mbox{$[t_j-T/2,t_j+T/2]$}.

\subsection{Coherent matched-filtering of one segment}

\subsubsection{The $\F$-statistic}

The likelihood ratio~$\Lambda_j$ for the $j$th segment,
deciding between the hypothesis of a signal $h(t)$ with amplitude parameters~$\Amp$ 
and phase parameters~$\vDoppler$, and no signal being present,
is written as~\cite{jks1},
\begin{equation}
  \ln \Lambda_j = (x|h)_j - \frac{1}{2} (h|h)_j \,,
  \label{e:loglikelihood1}
\end{equation}
where the following inner product has been used~\cite{jks1},
\begin{equation}
  (x|y)_j \equiv \frac{2}{S_n^{[j]}} \int_{t_j-T/2}^{t_j+T/2}\, x(t)\,y(t) \rmd t \,.
  \label{e:scalarpro2}
\end{equation}
with  $S_n^{[j]}$ defined as the one-sided noise
spectral density for the $j$th segment. 
Since this work is concerned with narrow-bandwidth
signals, $S_n^{[j]}$ is here taken as constant.

As was done in Ref.~\cite{jks1}, the following inner products are combined 
for every segment~$j$ into a $4\times4$ matrix~$\mathcal{M}^{[j]}$ 
whose components are
\begin{equation}
  \mathcal{M}_{\mu\nu}^{[j]} \equiv (h_\mu | h_\nu)_j \,,
  \label{e:Mmunu}
\end{equation}
where  \mbox{$\mu,\nu=1,2,3,4$}.
To very good accuracy, one can approximate~\cite{jks1},
\begin{subequations}
\begin{align}
 (h_1 | h_3)_j &\approx (h_1 | h_4)_j \approx (h_2 | h_3 )_j \approx (h_2 | h_4)_j \approx 0 \,,\\[1ex]
 (h_1 | h_1)_j &\approx (h_3 | h_3)_j \approx \frac{1}{2}A_j \,,\\[1ex]
 (h_2 | h_2)_j &\approx (h_4 | h_4)_j \approx \frac{1}{2}B_j \,,\\[1ex]
 (h_1 | h_2)_j &\approx (h_3 | h_4)_j \approx  \frac{1}{2} C_j\,,
\end{align}
\end{subequations}
with the definitions
\begin{equation}
A_j \equiv (a|a)_j  \,, \hspace{4mm} 
B_j \equiv (b|b)_j \,, \hspace{4mm} 
C_j \equiv (a|b)_j \,.
\label{e:ABC}
\end{equation}
In addition, we abbreviate the linear correlations $(x | h_\mu)_j$ by 
the following compact notation:
\begin{equation}
   \xx_\mu^{[j]} \equiv (x | h_\mu)_j\,.
\end{equation}
Thus,  \Eref{e:loglikelihood1} is rewritten as
\begin{equation}
\ln \Lambda_j =  \sum_{\mu=1}^{4} \Amp_\mu\, \xx_\mu^{[j]}
   - \frac{1}{2} \sum_{\mu,\nu=1}^{4}  \Amp_\mu\, \mathcal{M}_{\mu\nu}^{[j]}\,  \Amp_\nu  \,.
  \label{e:loglikelihood3}
\end{equation}

For every segment $j$ the log-likelihood ratio of \Eref{e:loglikelihood3}
is analytically maximized over the amplitude parameters $\Amp$.
The maximum likelihood (ML) estimators for~$\Amp$
obtained from the $j$th segment are denoted by
\begin{equation}
  {\hat \Amp}^{[j]} =\left(\hat\Amp_1^{[j]} ,\hat\Amp_2^{[j]} ,\hat\Amp_3^{[j]} ,\hat\Amp_4^{[j]} \right)\,,
\end{equation}
are explicitly given by~\cite{jks1},
\begin{align}
\label{e:MLEsSeg}
&\hat \Amp_{1}^{[j]} = 2\,\frac{ B_j\, \xx_1^{[j]} - C_j \, \xx_2^{[j]} }{D_j} \,,\qquad
\hat \Amp_{2}^{[j]}  = 2\,\frac{ A_j\, \xx_2^{[j]} - C_j \, \xx_1^{[j]} }{D_j} \,,
\nonumber\\[1ex]
&\hat \Amp_{3}^{[j]} = 2\,\frac{ B_j \, \xx_3^{[j]} - C_j \, \xx_4^{[j]} }{D_j} \,,\qquad
\hat \Amp_{4}^{[j]}  = 2\,\frac{ A_j \, \xx_4^{[j]} - C_j \, \xx_3^{[j]} }{D_j} \,,
\end{align}
where \mbox{$D_j \equiv A_j B_j - C_j^2$}, and $D_j\neq0$ has been assumed.

Replacing the amplitude parameters $\Amp$ in $\ln \Lambda_j$ 
of \Eref{e:loglikelihood3} with their ML estimators~$ {\hat \Amp}^{[j]} $
given by \Esref{e:MLEsSeg} yields the so-called \mbox{$\F$-statistic} 
for the $j$th segment,
\begin{align}
    \F_j &\equiv  \frac{B_j}{D_j} \left( {\xx_1^{[j]}}^2 + {\xx_3^{[j]}}^2 \right) 
    + \frac{A_j}{D_j} \left( {\xx_2^{[j]}}^2 + {\xx_4^{[j]}}^2 \right) \nonumber\\
    &\hspace{0.5cm}- \frac{2 C_j}{D_j}  \left( \xx_1^{[j]} \, \xx_2^{[j]} +  \xx_3^{[j]} \, \xx_4^{[j]} \right)  \,.
    \label{e:Fstat1}
\end{align}
This expression can be written compactly by using the four-vector notation
for the set of four linear correlations~$\xx_\mu^{[j]}$ as
\begin{equation}
   \xx^{[j]} \equiv (\xx_1^{[j]},\xx_2^{[j]},\xx_3^{[j]},\xx_4^{[j]}) \,,
   \label{e:xjvec}
\end{equation}
such that \Eref{e:Fstat1} takes the form
\begin{align}
      \F_j = \frac{1}{2}\,  \xx^{[j]} \; {\mathcal{M}^{[j]}}^{-1} \; {\xx^{[j]}}^\trapo \,,
    \label{e:Fstat2}
\end{align}
where the superscript $^\trapo$ indicates the transpose. 
Therefore, the $\F$-statistic represents a quadratic form in terms of the
linear correlations~$\xx_\mu^{[j]}$. 
It should be noted that in practice~$\F_j$ can be efficiently computed using the FFT algorithm
when rewriting the four linear correlations $\xx_\mu^{[j]}$ as two complex integrals;
further details are described in~\cite{jks1,Resamp2010}.

We find that the $\F$-statistic can be equivalently formulated 
as a quadratic form in terms of the ML estimators~${\hat \Amp_\mu}^{[j]}$.
Using \Esref{e:MLEsSeg} to substitute the~$\xx_\mu^{[j]}$ in \Eref{e:Fstat1} yields
\begin{align}
 \F_j &= \frac{A_j}{4} \left(  {\hbox{$\hat \Amp_1$}^{[j]}}^2 + {\hbox{$\hat \Amp_3$}^{[j]}}^2 \right)
 + \frac{B_j}{4} \left( {\hbox{$\hat \Amp_2$}^{[j]}}^2 + {\hbox{$\hat \Amp_4$}^{[j]}}^2 \right) \nonumber\\
 &\hspace{0.5cm}+ \frac{C_j}{2} \left(\hat \Amp_{1} \hat\Amp_2 +  \hat \Amp_{3}  \hat\Amp_4 \right),
 \label{e:Fstat1mle}
\end{align}
which is compactly rewritten as
\begin{equation}
      \F_j = \frac{1}{2}\, {\hat \Amp}^{[j]} \, \mathcal{M}^{[j]} \, {\hbox{$\hat \Amp$}^{[j]}}^\trapo \,,
    \label{e:Fstat2mle}
\end{equation}
showing that the $\F$-statistic can also be viewed as quadratic form in terms of 
the ${\hat \Amp_\mu}^{[j]}$ with a coefficient matrix being equal to~$\mathcal{M}^{[j]}$.
This formulation \eref{e:Fstat2mle} of the $\F$-statistic is not common in the existing literature, 
but closely related is the work of~\cite{PrixWhelan2007,WPK2008,WPK2010}. There, the matrix
$\mathcal{M}^{[j]}$ is considered  as a metric on the amplitude parameter space~$\Amp$,
and a norm of the four-vector~$\Amp$ is defined by
\mbox{$ ||\Amp|| \equiv \sqrt{\Amp \, \mathcal{M}^{[j]} \, \Amp^\trapo }$}.
In this context, we see from \Eref{e:Fstat2mle} that the \mbox{$\F$-statistic} is simply
half the squared ``length'' of the four-vector~${\hat{\Amp}}^{[j]}$ of amplitude ML estimators:
\mbox{$\F_j =  ||{\hat{\Amp}}^{[j]}||^2/2$}.

\subsubsection{Statistical properties}

The four linear  correlations $\xx_\mu^{[j]}$ for a given segment~$j$ are Gaussian
distributed random variables, whose
expectation values and variances, respectively, are 
in absence of a signal, when \mbox{$x(t)=n(t)$}, obtained as 
\begin{equation}
 E_n\left[ \xx_\mu^{[j]} \right] =0\,,\qquad
 E_n\left[ \xx_\mu^{[j]} \,  \xx_\nu^{[j]} \right]  =  \mathcal{M}_{\mu\nu}^{[j]}\,. 
 \label{e:Expxxn}
\end{equation}
Therefore, in this case, the probability density function of $2\F_j$ 
is a central $\chi^2$~distribution with $4$ degrees of freedom~\cite{jks1}.
Hence, $2\F_j$ has the following expectation value and variance, respectively:
\begin{equation}
  E_n\left[ 2\F_j \right]=4\,,\qquad  \sigma_{2\F_j,n}^2=8 \,.
\end{equation}

When a signal is present, which perfectly matches the
template waveform $h(t)$, then the expectation values 
corresponding to \Esref{e:Expxxn} are obtained as  
\begin{align}
   E_h\left[ \xx_\mu^{[j]} \right] &=(h|h_\mu)_j\,,\\
   E_h\left[ \xx_\mu^{[j]} \,  \xx_\nu^{[j]} \right]  &=  \mathcal{M}_{\mu\nu}^{[j]} + (h|h_\mu)_j\,(h|h_\nu)_j\,.
\end{align}
Thus, as first noted in~\cite{jks1}, the covariance matrix for the Gaussian random
variables $\xx_\mu^{[j]}$ is the \emph{same} whether a signal is
present or not, and it is exactly equal to $\mathcal{M}^{[j]}$.
It should also be noted that the inverse of $\mathcal{M}^{[j]}$ 
is equal to the covariance
matrix of the ML estimators~${\hat \Amp_\mu}^{[j]}$.
Thus, in this case, $2\F_j$ has noncentral $\chi^2$ distribution
with $4$ degrees of freedom and a noncentrality parameter~$\rho_j^2\equiv(h|h)_j$,
where $\rho_j$ is commonly referred to as 
the ``optimal'' signal-to-noise ratio (S/N). Thus, the expectation value 
and variance of $2\F_j$ in this perfect-match case are  
\begin{equation}
  E_h\left[ 2\F_j \right]=4+\rho_j^2\,,\qquad
  \sigma_{2\F_j,h}^2=8+4\rho_j^2\,,
\end{equation}
where $\rho_j^2$ is explicitly obtained as
\begin{align}
  \rho_j^2 &=  A_j\, \frac{ \Amp_{1}^2 + \Amp_{3}^2}{2}
 + B_j \, \frac{ \Amp_{2}^2 +  \Amp_{4}^2}{2}
 + C_j \left(\Amp_{1} \Amp_2 +  \Amp_{3}  \Amp_4 \right) \nonumber\\
 &= \Amp \, \mathcal{M}^{[j]} \, {\Amp}^\trapo \,,
 \label{e:optSNR}
\end{align}
with $\Amp$ representing the $4$-vector of the signal's amplitude 
parameters, \mbox{$\Amp=(\Amp_1,\Amp_2,\Amp_3,\Amp_4)$}.
Comparing \Eref{e:optSNR} to \Eref{e:Fstat2mle}, we find
that twice the $\F$-statistic can  be interpreted as the ML estimator
for the squared S/N: \mbox{$2\F_j = \hat{\rho}^2_j$}.

\subsection{Incoherent combination of coherently analyzed segments}

\subsubsection{The standard hierarchical detection statistic \label{ssec:StanSemiCohDetStat}}

We denote the standard hierarchical detection statistic by~$\bar{\F}$,
which, as used in~\cite{cutler:2005,pletsch:GCT,pletsch:scmetric},
represents the sum of one $\F$-statistic value $\F_j$ from each segment~$j$, 
\begin{equation}
  \bar{\F} =  \sum_{j=1}^N \F_j = \frac{1}{2}\,\sum_{j=1}^N  \xx^{[j]} \; {\mathcal{M}^{[j]}}^{-1} \; {\xx^{[j]}}^\trapo  \,,
  \label{e:barF1}
\end{equation}
evaluated at a given fine-grid point in phase parameter space.
Therefore, $\bar{\F}$ also represents a quadratic form in terms of the linear correlations
$\xx_\mu^{[j]}$, and one can compactly rewrite \Eref{e:barF1} as
\begin{equation}
   \bar \F = \frac{1}{2}\, \xx \, \bar{\mathcal{M}}^{-1} \, \xx^\trapo \,,
   \label{e:barFc}
\end{equation}
where the $4N$-vector $\xx$ collects all the $\xx^{[j]}$ as
\begin{equation}
    \xx = (\, \xx^{[1]},\xx^{[2]},...,\xx^{[N]} \,) \,.
    \label{e:4Nxx}
\end{equation}
and the $4N\times4N$ matrix $\bar{\mathcal{M}}$ is defined to have the form
\begin{equation}
 \bar{\mathcal{M}} = \begin{pmatrix}
  \mathcal{M}^{[1]} &   & &\smallskip\\
    &  \mathcal{M}^{[2]} & &\smallskip\\
        &   &\ddots &\smallskip\\
            &   & & \mathcal{M}^{[N]}\smallskip\\
\end{pmatrix} \,.
\end{equation}
Hence, the standard hierarchical scheme, using the detection statistic~$\bar\F$, represents
the incoherent combination of \emph{epochwise} coherent matched-filter outputs, 
where each epoch has duration~$T$.
Recall that, using \Eref{e:Fstat2mle}, $\bar{\F}$ can equivalently be 
rewritten as a quadratic form in terms of the amplitude ML estimators ${\hat\Amp}^{[j]}$,
leading to
\begin{equation}
  \bar{\F} = \frac{1}{2}\,\sum_{j=1}^N  {\hat\Amp}^{[j]} \; {\mathcal{M}^{[j]}} \; {\hbox{$\hat \Amp$}^{[j]}}^\trapo
  =\frac{1}{2}\,{\hat\Amp} \; \bar{\mathcal{M}} \; {\hat \Amp}^\trapo
  \,,
  \label{e:barF2}
\end{equation}
where $\hat\Amp$ denotes the $4N$-vector \mbox{$\hat \Amp = (\,{\hat\Amp}^{[1]},{\hat\Amp}^{[2]},...,{\hat\Amp}^{[N]}\,)$}.

The problem of efficiently selecting the best coarse-grid $\F_j$ value in every
segment for a given fine-grid point has been studied in previous work~\cite{pletsch:GCT,pletsch:scmetric},
and hence for rest of this paper it is assumed that such an efficient link
between the coarse and fine grids is available.

\subsubsection{Statistical properties}

In absence of a signal, it is straightforward to show that the probability density function of $2\bar\F$ 
is a central $\chi^2$ distribution with $4N$ degrees of freedom~\cite{jk3,cutler:2005};
hence, $\bar\F$ has the expectation value and variance, respectively,
\begin{equation}
  E_n\left[ 2\bar\F \right]=4N\,,\qquad 
  \sigma_{2\bar\F,n}^2=8N \,.
  \label{e:ExpVarFbar-n}
\end{equation}
On the other hand, if a signal $h(t)$ is present, 
perfectly matching the template-waveform phase parameters, then
$\bar\F$ has a noncentral $\chi^2$ distribution with $4N$ degrees of freedom
and noncentrality parameter~${\bar\rho}^2$.
The expectation value and variance of $\bar\F$, respectively, are obtained as 
\begin{equation}
  E_h\left[ 2\bar\F \right]=4N+\bar\rho^2\,,\qquad
  \sigma_{2\bar\F,h}^2=8N+4\bar\rho^2\,,
  \label{e:ExpVarFbar-h}
\end{equation}
where $\bar\rho$ is given by
\begin{equation}
   \bar\rho^2 \equiv \sum_{j=1}^N \rho_j^2 \,,
   \label{e:rhobar1}
\end{equation}
recalling that $\rho_j$ as of  \Eref{e:optSNR} denotes the optimal S/N  for the $j$th segment.

It is interesting to note that $\bar\rho$ of \Eref{e:rhobar1} is 
actually \emph{equal} to the fully coherent optimal S/N
for the entire data set. The apparent difference in search sensitivity
results from the different underlying probability distributions.
In the standard hierarchical scheme, there are $N$ times as
many degrees of freedom as compared to the fully coherent case.

\section{Sliding coherence window approach\label{sec:SlideWin}}

\begin{figure}[t]
	\includegraphics[width=0.85\columnwidth]{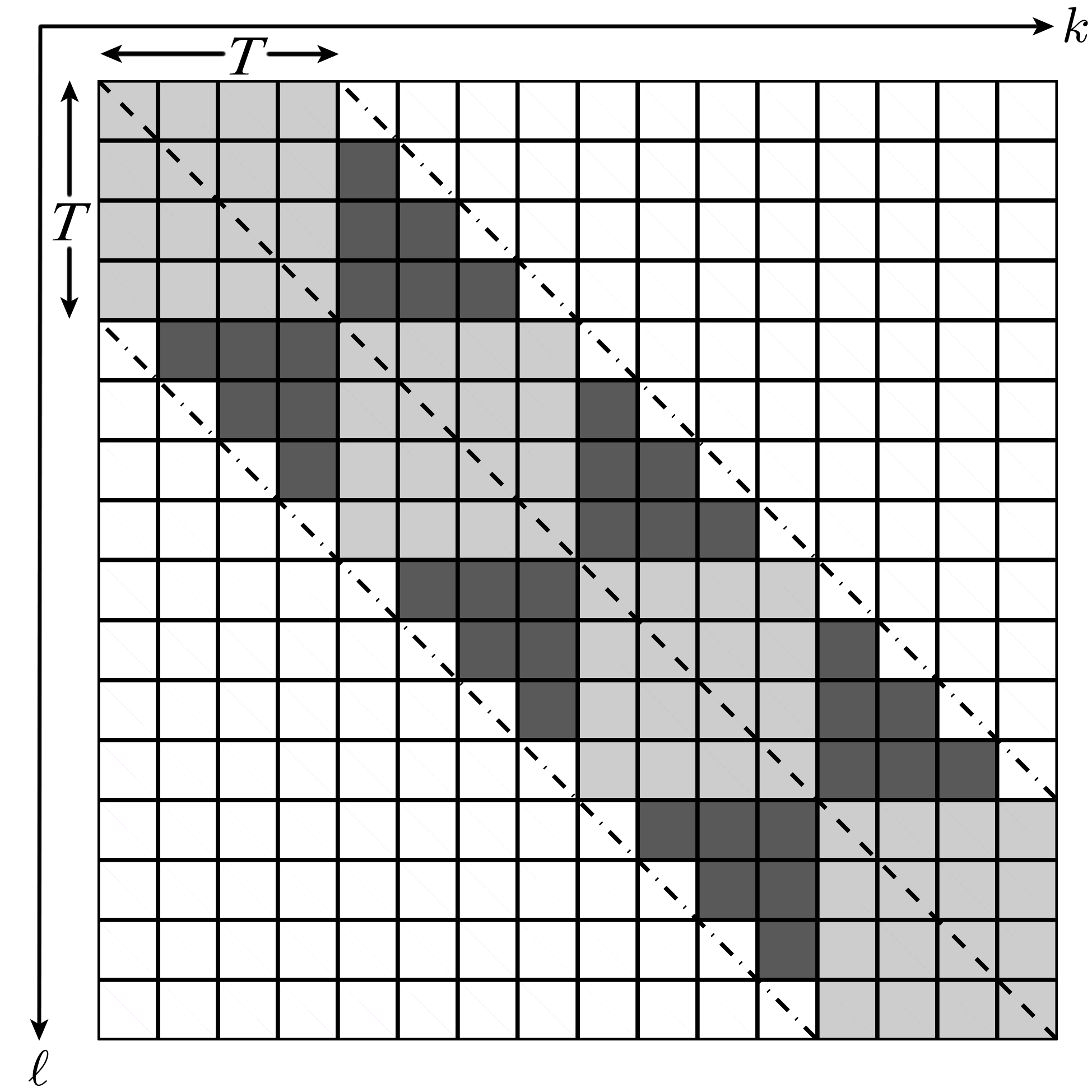}
	\caption{Schematic comparison of the standard hierarchical search scheme
	and the sliding coherence window approach for the same
	coherence time baseline~$T$. Each box represents one product of the 
	linear correlations \mbox{$\xx_\mu^{[k]} \xx_\mu^{[\ell]}$}, obtained from the
	subsegments~$k$ and~$\ell$, respectively.
	In this example, the indices take the values $k,\ell=1,...,16$.
	The light grey boxes represent the set of products selected in the 
	standard hierarchical search scheme, for~\mbox{$N=4$} coherent segments 
	of duration~$T$. 
	The dark grey boxes are the products \emph{additionally} selected 
	by the sliding coherence window technique to enhance the search sensitivity.
	In this example, $T$ is subdivided into~\mbox{$q=4$} subsegments 
	(implying a coherence overlap of $\coho=75\%$ between successive sliding steps).
	In contrast, a fully coherent search over the entire data set would have to 
	include all boxes shown.
	\label{f:MLEsCorr}}
\end{figure}

The central idea behind the sliding coherence window scheme is to
use a window the size of the coherence time baseline~$T$ and to
``slide'' it over the data set in steps \emph{smaller} than~$T$ to combine
the coherent matched-filter outputs from each sliding step. 
This effectively amounts to the incoherent combination of coherent matched-filter outputs
from \emph{overlapping} segments of length~$T$,
which remarkably improves the search sensitivity compared to the
standard hierarchical scheme. 
There, only coherent matched-filter outputs from \emph{nonoverlapping} segments 
of coherence length~$T$ are combined, omitting to coherently correlate
large parts of the data which still lie \emph{within} the coherence time baseline~$T$,
as illustrated in \Fref{f:MLEsCorr}. 

The number of templates to discretely cover the phase parameter space
searched is of great importance, since this is what ultimately limits
the overall search sensitivity at the finite computing power available.
Therefore, it should also be emphasized that the sliding coherence window 
technique employs the \emph{same} number of templates
in phase parameter space as the standard scheme.
The combination of overlapping coherent integrations alters neither
the coherence time baseline~$T$ nor the total data time span $T_{\textrm{\tiny data}}$.
Hence, it is obvious that the same semicoherent metric as 
previously studied in~\cite{pletsch:GCT,pletsch:scmetric} can also be used
in combination with the here-proposed sliding coherence window technique~\footnote{One
may further illustrate this by noting that the semicoherent metric is effectively the average of 
all individual-segment coherent metrics \cite{bc2:2000,pletsch:GCT,pletsch:scmetric}.
Thus, in this sense, at fixed coherence baseline $T$ and at fixed total data time span
$T_{\textrm{\tiny data}}$, the additional overlapping segments do not change 
the averaged metric~\cite{privcomm1}.}.

\subsection{Detection statistic\label{ssec:SlideWinDetStat}}

For computational efficiency, the sliding coherence window approach subdivides
every data segment of duration~$T$ into~$q$ ``subsegments," as 
adumbrated in \Fref{f:MLEsCorr}.
Hence, each subsegment is of duration $T_q=T/q$, which represents
the step size between each sliding iteration of the coherence window.

The subsegments are labeled by \mbox{$k=1,...,qN$}.
Thus, in analogy to \Eref{e:xjvec}, we define the four-vector $\xx^{[k]}$
for the $k$th subsegment as
\mbox{$\xx^{[k]} =(\,\xx_1^{[k]},\xx_2^{[k]},\xx_3^{[k]},\xx_4^{[k]} \,)$}.
In this context, similarly to \Eref{e:4Nxx}, the $4qN$-vector now collects all the $\xx^{[k]}$,
\mbox{$\xx = (\xx^{[1]},\xx^{[2]},...,\xx^{[qN]})$}.
Analogously,  the same notation also applies to the $4qN$ amplitude 
ML estimators~${\hat \Amp}_\mu^{[k]}$.

As sketched in \Fref{f:MLEsCorr}, the central goal of the sliding coherence window strategy 
is to combine a larger number of \emph{distinct} pairs~$\{\xx_\mu^{[k]},\xx_\nu^{[\ell]}\}$~\footnote{
Equivalently, one could also employ the 
pairs \mbox{$\{\hat{\Amp}_\mu^{[k]},\hat{\Amp}_\nu^{[\ell]}\}$}.
},
while still restricting the maximum difference between their timestamps
$t_k$ and $t_\ell$ to at most the coherent time baseline~$T$.
Hence, the resulting selection condition is \mbox{$|t_k - t_\ell| \leq T$}.

To achieve this goal, in principle an appropriate $4qN\times4qN$ 
coefficient matrix $\mathcal{U}$ needs to be constructed, 
constituting the following quadratic form~$\Astat$,
\begin{equation}
   \Astat = \frac{1}{2}\,  \xx \;{\mathcal{U}} \; \xx^\trapo \,,
   \label{e:Astat1}
\end{equation}
which represents the detection statistic 
of the sliding coherence window search technique.

In order to simplify the construction of $\mathcal{U}$, 
we exploit the fact that the constants $C_k$ are typically much smaller than 
the values of $A_k$ and $B_k$, and therefore terms involving $C_k$ 
are neglected~\footnote{
The covariance matrix of the Gaussian random variables~$\xx_\mu^{[k]}$
can always be diagonalized via a linear transformation given by Eq.(65) of Ref.~\cite{jks1}.
Besides, $C_k$ exactly vanishes when computed over a time interval of a multiple 
of one sidereal day~\cite{jk3}. 
Also note that when averaged over all sky positions $(\alpha,\delta)$, 
    in contrast to $A_k$ and $B_k$, the constant $C_k$ vanishes.
}. 
With this approximation, \Eref{e:Astat1} should explicitly read as
\begin{align}
 \Astat &=  \sum_{k,\ell=1}^{qN}  Q_{T}(t_k-t_\ell)
  \Biggl[ \frac{ {\xx_1^{[k]}} {\xx_1^{[\ell]}} + {\xx_3^{[k]}} {\xx_3^{[\ell]}} }{\sqrt{A_k\,A_\ell}} 
  +  \frac{ {\xx_2^{[k]}} {\xx_2^{[\ell]}} + {\xx_4^{[k]}} {\xx_4^{[\ell]}} }{\sqrt{B_k\,B_\ell} }
\Biggr],
    \label{e:Astat2}
\end{align}
where the step function $Q_{T}(x)$ selects the pairs of linear correlations 
according to their time difference and the predefined coherent time baseline~$T$,
\begin{equation}
   Q_{T}(x) \equiv \begin{cases} 1 & |x| \leq T \\ 0 & |x| > T\end{cases} \,.
   \label{e:Qstep}
\end{equation}
As with $\bar\F$, the detection statistic~$\Astat$ can also be equivalently reformulated
as a quadratic form in terms of the amplitude ML estimators~${\hat \Amp}_\mu^{[k]}$.
For practical convenience, in what follows we use $\xx_\mu^{[k]}$
as in \Eref{e:Astat2}. However, note that in the above approximative case,
\mbox{$\xx_\mu^{[k]} \propto {\hat \Amp}_\mu^{[k]}$},
thus making the interchange 
between $\xx_\mu^{[k]}$ and ${\hat \Amp}_\mu^{[k]}$ simple if desired.

When \mbox{$q=1$}, it is obvious that $\Astat$ coincides with the standard
hierarchical detection statistic $\bar\F$. 
However, if one chooses \mbox{$q>1$},
the detection statistic~$\Astat$ is able to improve performance compared 
to~$\bar\F$, as will be described in what follows.

Moreover, a useful quantity is denoted by~$\coho$, which defines
the average ``coherence overlap'' between successive sliding steps.
For the case of a contiguous data set~\footnote{
When the data contains gaps, the average coherence overlap~$\coho$ is a function of
the subsegment time midpoints~$t_k$.}, as considered in this presentation, 
$\coho$ is related to~$q$ simply via \mbox{$ \coho = 1 - 1/q$}.

\subsection{Statistical properties and sensitivity estimation\label{ssec:StatsAndSenEst}}

To analytically estimate the sensitivity of the sliding coherence window 
search, the underlying statistical properties are examined.
Recall that, for simplicity, the data set has been taken as free of gaps, such that one can
write the time span of the entire data set as \mbox{$\Tdata=NT=qN\,T_q$}.
The detection statistic $\Astat$ of \Eref{e:Astat2} is explicitly written as
\begin{align}
   \Astat &=  \sum_{k=1}^{qN}  \Biggl\{ 
   \frac{ {\xx_1^{[k]}}^2 + {\xx_3^{[k]}}^2 }{A_k}  + \frac{ {\xx_2^{[k]}}^2 + {\xx_4^{[k]}}^2 }{B_k} \nonumber\\
   &+ 2\sum_{\ell=k+1}^{k+q-1} \Biggl[ 
   \frac{ {\xx_1^{[k]}} {\xx_1^{[\ell]}} + {\xx_3^{[k]}} {\xx_3^{[\ell]}} } { \sqrt{A_k A_\ell} } 
   + \frac{ {\xx_2^{[k]}} {\xx_2^{[\ell]}} + {\xx_4^{[k]}} {\xx_4^{[\ell]}} } { \sqrt{B_k B_\ell} } 
   \Biggr]\Biggr\} \,.
    \label{e:Astat3}
\end{align}
Since the actual probability density function of~$\Astat$ is cumbersome to work
with, we approximate it here by a Gaussian distribution, 
which is well justified based on the generalized central limit theorem (provided \mbox{$N\gg1$}),
as done similarly in previous work~\cite{hough:2005}.
Thus, we proceed by computing the mean and variance of~$\Astat$.

When the data consist of zero-mean stationary Gaussian noise only, 
the expectation value of $2\Astat$ is obtained as
\begin{equation}
 E_n\left[2 \Astat \right] = 4qN \,,
 \label{e:EnAstat}
\end{equation}
and the variance of $2\Astat$ is given by
\begin{equation}
\label{e:sigmasqAstatn}
 \sigma_{2\Astat,n}^2 =  8 qN(2q-1)  \,.
\end{equation}
It is straightforward to show that for Gaussian noise, a certain
false alarm probability $P_{\textrm{FA}}$  corresponds to a 
threshold $\Astat_{\textrm{th}}$ via
\begin{equation}
 \Astat_{\textrm{th}} 
  =   E_n\left[ \Astat \right] + \sigma_{\Astat,n}\, \sqrt{2} \; \erfc^{-1}\left(2 P_{\textrm{FA}} \right)\,,
  \label{e:Astatth}
\end{equation}
where $\erfc$ denotes the complementary error function.

Provided the presence of a signal~$h(t)$
whose phase parameters perfectly match 
the template, then the expectation value of $2\Astat$ is given by
\begin{equation}
  E_h\left[ 2\Astat \right] = 4qN +  \rho_{\Astat}^2 \,,
   \label{e:Exp2Z-h}
\end{equation} 
where we defined $\rho_{\Astat}$ as
\begin{align}
 \rho_{\Astat}^2 &\equiv \frac{\Amp_1^2+\Amp_3^2}{2} \left\{  \sum_{k=1}^{qN} \left[ 
  A_k  + 2\sum_{\ell=k+1}^{k+q-1}     \sqrt{A_k A_\ell}  \right] \right\} \nonumber\\
  &\hspace{0.45cm}+\frac{ \Amp_2^2+\Amp_4^2 }{2} \left\{  \sum_{k=1}^{qN} \left[  
  B_k + 2\sum_{\ell=k+1}^{k+q-1}     \sqrt{B_k B_\ell}  \right] \right\}.
  \label{e:SNRsqAstat}
\end{align}
The probability of detection $P_{\textrm{DET}}$ for Gaussian noise is given by
\begin{equation}
  P_{\textrm{DET}} 
  = \frac{1}{2} \;\erfc\left( \frac{\Astat_{\textrm{th}} - E_h\left[\Astat \right]}{\sqrt{2}\; \sigma_{\Astat,h}} \right) \,.
  \label{e:PDET}
\end{equation}
For current ground-based detectors, the expected CW signals are extremely weak,
so that the small-signal situation (\mbox{$h\ll n$}) is well justified.
Thus, we approximate $\sigma_{\Astat,h}$ by using 
$\sigma_{\Astat,n}$  and, by means of \Esref{e:EnAstat} - \eref{e:Exp2Z-h}, 
one obtains from \Eref{e:PDET} the following relation:
\begin{equation}
   \rho_{\Astat}^2 = \statfac \sqrt{2}\, \sigma_{\Astat,n} \,,
   \label{e:SNRsqAstat2}
\end{equation}
where $\statfac$ has been defined as
\begin{equation}
  \statfac \equiv \erfc^{-1}\left(2 P_{\textrm{FA}}  \right) - \erfc^{-1}\left(2 P_{\textrm{DET}}  \right) \,.
  \label{e:statfac}
\end{equation}
The minimum detectable gravitational-wave strain tensor amplitude $h_0$
can be determined from \Eref{e:SNRsqAstat2}, 
because \mbox{$h_0^2\propto \rho_{\Astat}^2$}  as follows from \Eref{e:SNRsqAstat}.

To obtain the estimated sensitivity scaling of the sliding coherence window search
in terms of the most relevant parameters, the noise floor $S_n$ is taken as constant 
throughout the data set.
In addition, we replace the constants $A_k$ and $B_k$ by effective average values as
\mbox{$A_k\approx \frac{2T_q}{ S_n} \,\bar A$}, and \mbox{$B_k\approx \frac{2T_q}{ S_n} \,\bar B$},
and define $\bar\kappa$ as
\begin{equation}
  \bar\kappa\equiv \bar{A}\,(\bar{\Amp_1}^2 + \bar{\Amp_3}^2) + \bar{B}\,(\bar{\Amp_2}^2 + \bar{\Amp_4}^2)\,, 
\end{equation}
where the $\bar\Amp_\mu$ are the same as the $\Amp_\mu$ apart from the factor~$h_0$,
\mbox{$\bar\Amp_\mu \equiv \Amp_\mu/h_0$}.
Thus, \Eref{e:SNRsqAstat} simplifies to
\begin{equation}
  \rho_{\Astat}^2  = h_0^2\, \bar\kappa \frac{T}{S_n} N\left(2q-1\right)\,.
  \label{e:rhoAstatSq}
\end{equation}
In turn, using \Eref{e:rhoAstatSq} to substitute $\rho_{\Astat}^2$ in \Eref{e:SNRsqAstat2}
and solving for $h_0$ yields
\begin{equation}
  h_0 = \frac{2\sqrt{\statfac}}{\sqrt{\bar\kappa}} \sqrt{\frac{S_n}{T}} \;N^{-1/4} \left(2-\frac{1}{q}\right)^{-1/4} \,,
  \label{e:h0SLW}
\end{equation}
revealing the estimated sensitivity scaling
of the sliding coherence window search.  One may further rewrite \Eref{e:h0SLW}  as
\begin{equation}
  h_0 = \frac{2\sqrt{\statfac}}{\sqrt{\bar\kappa}} \sqrt{\frac{S_n}{T}} \;
  \left[N\, \left(1+\coho\right)\right]^{-1/4} \,,
  \label{e:h0SLWcoho}
\end{equation}
using the previously introduced average coherence overlap~$\coho$.

\subsection{Comparison of sensitivity with standard scheme\label{ssec:cmpSen}}

\begin{figure}[t]
	\includegraphics[width=0.93\columnwidth]{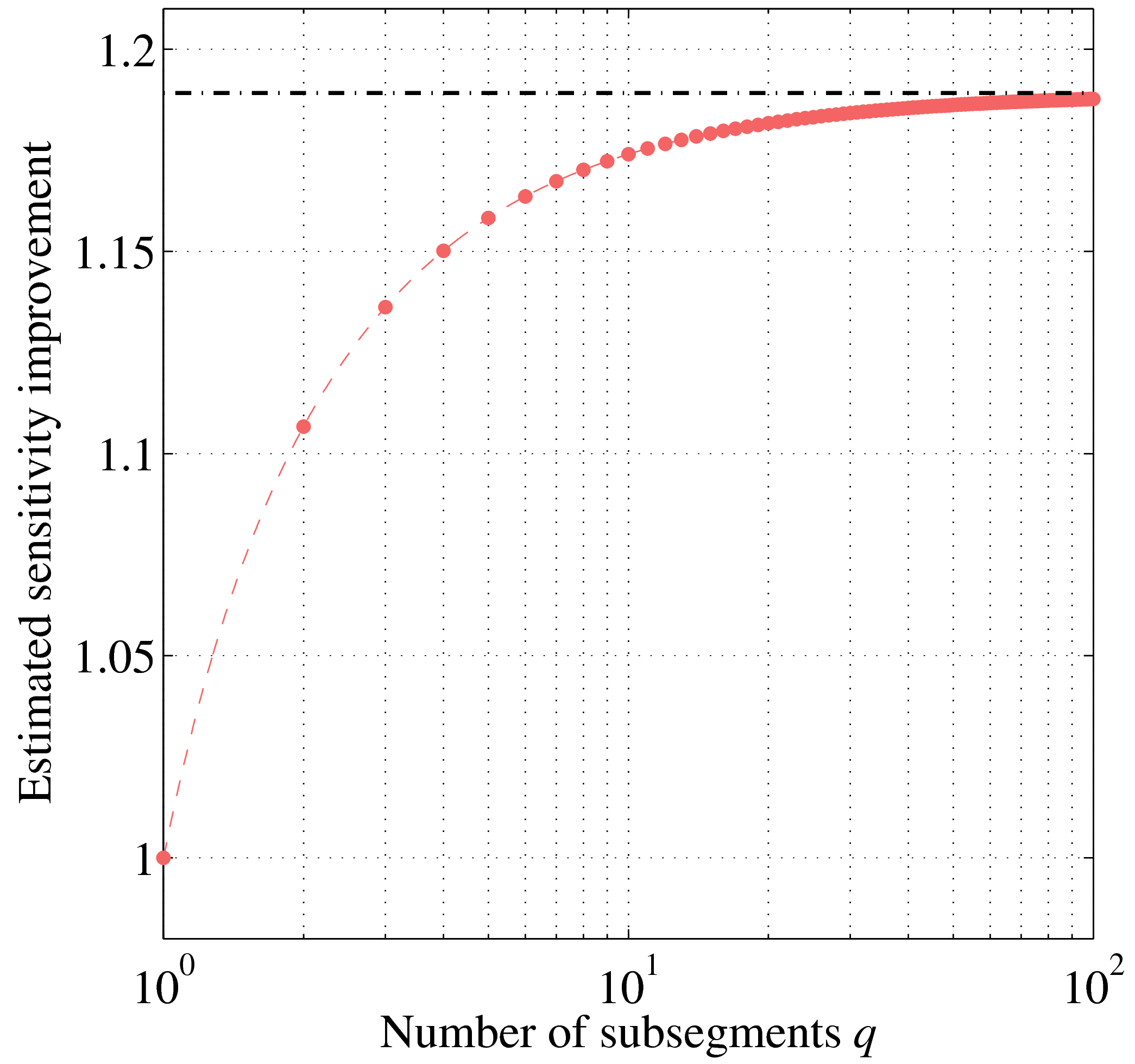}
	\caption{Estimated sensitivity improvement factor of the sliding coherence 
	window technique over the standard hierarchical search strategy,
	shown as a function of~$q$ (number of subsegments).
	The curve is explicitly given by \mbox{$(2-1/q)^{1/4}$}. 
	The horizontal dashed-dotted line indicates the constant value $2^{1/4}\approx1.19$.
	\label{f:senimprfac}}
\end{figure}

Equation~\eref{e:h0SLWcoho} also reveals the estimated 
sensitivity improvement of the sliding coherence window technique
compared to the standard hierarchical search under the same assumptions.
The standard hierarchical scheme is recovered for $\coho=0$ (i.e., $q=1$). 
Therefore, the sliding coherence window approach is more sensitive
than the standard hierarchical search scheme by the factor
$(1+\coho)^{1/4}$. In terms of~$q$, the sensitivity improvement factor is
$(2-1/q)^{1/4}$, which is shown in~\Fref{f:senimprfac}.

The coherence overlap~$\coho$ enhances the search sensitivity
effectively as if increasing the number of segments in the standard method.
In other words, to achieve the same sensitivity as with the sliding coherence 
window technique at given~$T$, using the standard hierarchical search method, 
effectively \mbox{$\coho=1+1/q$}  more segments have to be analyzed
(hence $50-100\,\%$ more data).

 
In practice, the choice of~$q$ (or equivalently~$\coho$) will generally have to be 
optimized in terms of search sensitivity at the given computational constraints 
and code implementation at hand, as well as for the detector data available. 
Further investigation in this direction will be presented in
\Sref{sec:ComputingCost}, comparing the estimated search sensitivity 
at fixed computational cost.

\section{Sensitivity performance demonstration\label{sec:PerfDemo}}

The performance improvement of the sliding coherence window technique 
is illustrated through realistic Monte Carlo simulations. In particular,
receiver operating characteristic (ROC) curves are obtained
to compare the standard hierarchical search scheme that is \mbox{$q=1$}
(corresponding \mbox{to $\coho=0$})
and the sliding coherence window strategy 
for \mbox{$q=2$} (corresponding to $\coho=50\%$).

The simulated data set refers to the two \mbox{LIGO 4-km} detectors (H1 and L1)
and spans a time interval of $5\,000\,\hrs$.
To provide a realistic comparison, a typical value is taken for the coherent time 
baseline of \mbox{$T=50\,\hrs$}, which results in $N=100$.
The software tools used are part of LALApps~\cite{LALApps} and employ 
accurate barycentering routines with timing errors below~$4\mu$s~\cite{LigoS1pulpaper}. 

In this study, the phase parameter space considered is four dimensional
using one spindown parameter, as in current all-sky surveys for prior unknown 
CW sources~\cite{S4EAH,powerflux:2009,S5R1EAH}.
Thus a point in phase parameter space is labeled by \mbox{$\vDoppler=(f,\dot f,\alpha,\delta)$}.

The false alarm probabilities are found from thousands of different
realizations of stationary Gaussian white noise with 
\mbox{$\sqrt{S_n} =  3.25\times10^{-22}\,{\rm Hz}^{-1/2}$}. 
To obtain the detection probabilities, distinct CW signals with fixed
gravitational-wave strain tensor amplitude of \mbox{$h_0=1.0\times 10^{-24}$}
are added. The remaining parameters of the signal population are randomly
drawn from uniform distributions in~$\psi$, $\cos\iota$, $\Phi_0$,
in the entire sky, frequencies in the interval \mbox{$f\in [155.12,155.16]\,\Hz$},
and spindowns over the range of \mbox{$\dot f \in [-2.64, 0.264]\,\textrm{n}\Hz/\textrm{s}$}.

\begin{figure}[t]
	\includegraphics[width=\columnwidth]{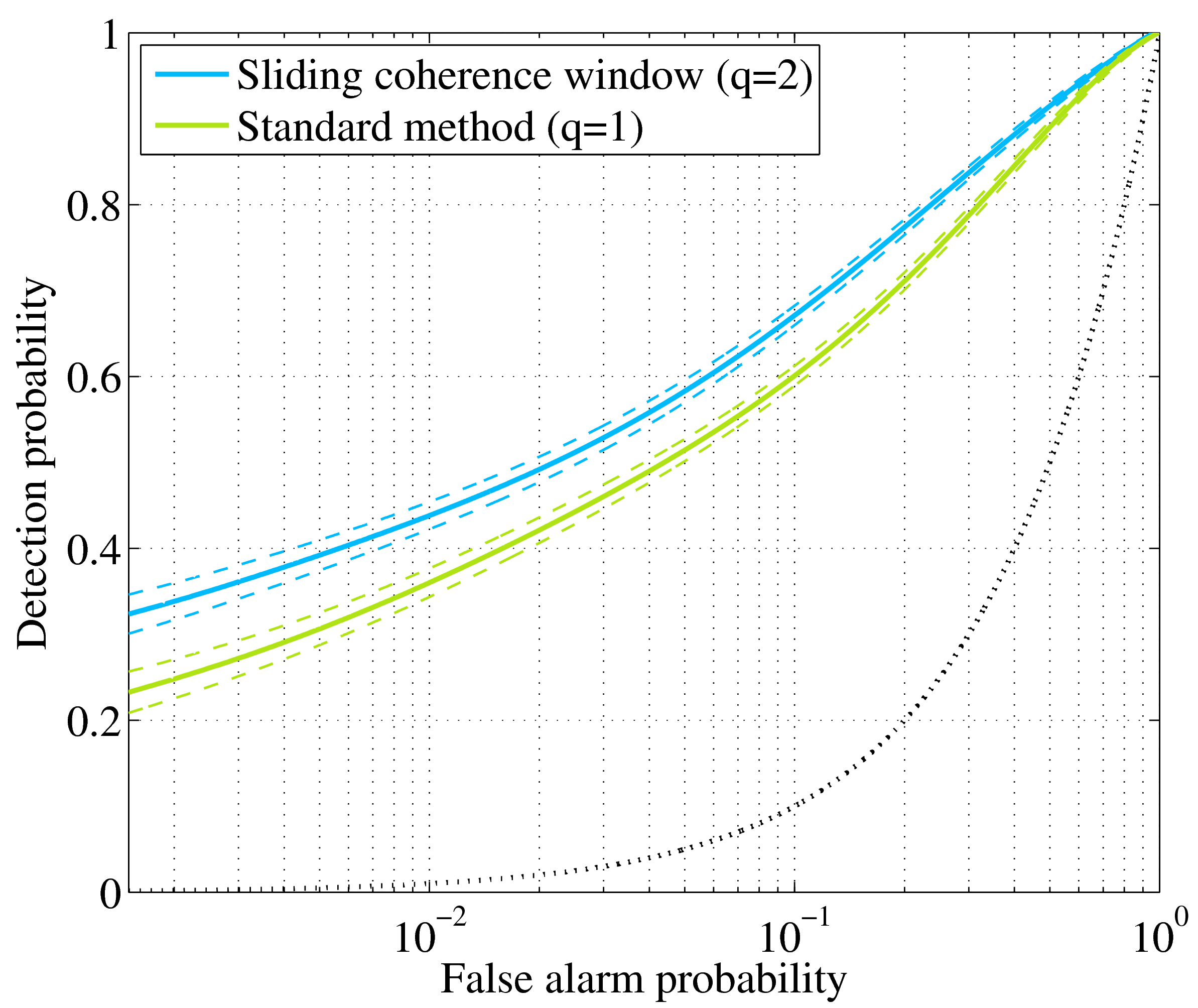}
	\caption{ROC curves comparing at fixed gravitational-wave
	amplitude~$h_0$ the standard hierarchical search technique 
	(lower solid curve) and the sliding coherence window method (upper solid curve) 
	with \mbox{$q=2$} implying a coherence overlap of $\coho=50\%$.
	The dashed curves on either side of the solid curves represent estimated $1\sigma$~errors. 
	The black dotted curve is the so-called line of no discrimination.
	\label{f:roc}}
\end{figure}

Figure~\ref{f:roc} compares the resulting ROC curves
for the different search techniques.
The ROC curves are computed from $6\,000$ different realizations.
The $1\sigma$~errors shown in \Fref{f:roc} are
based on a jackknife estimate as in~\cite{PrixKrishnanBstat2009,ConwaySloane1984} 
using $100$ subsets.
As expected,  the sliding coherence window technique with \mbox{$q=2$} is 
substantially more ``powerful'' than the standard scheme
(retrieved for \mbox{$q=1$}), yielding a higher
probability of detection  for the same false alarm probability.

Furthermore, the numerical results in \Fref{f:roc} attest to the analytically 
estimated gain in sensitivity obtained in \Eref{e:h0SLW}. For example,
at fixed false alarm probability of $P_{\textrm{FA}}=1\%$, the achieved
detection probability of the sliding coherence window technique 
is \mbox{$P_{\textrm{DET}}=0.427$}, whereas the standard hierarchical scheme
gives \mbox{$P_{\textrm{DET}}=0.356$}. To compare these values with
the theoretical expectation, note that \Eref{e:h0SLW} yields
\mbox{$\sqrt{\statfac} \propto (2-1/q)^{1/4}$}, where $\statfac$ has been
defined in \Eref{e:statfac} and solely depends on $P_{\textrm{FA}}$ 
and $P_{\textrm{DET}}$. Thus, for \mbox{$q=2$}, the predicted increase 
(compared to \mbox{$q=1$}) in $\sqrt{\statfac}$ is \mbox{$(3/2)^{1/4}\approx 10.7\%$}.
The above values obtained from the numerical simulations of \Fref{f:roc} 
yield a corresponding increase in $\sqrt{\statfac}$ of about $9.7\%$,
which is in agreement with the theoretical prediction
at the subpercent level.


\section{Comparison at fixed computing cost\label{sec:ComputingCost}}

In \Sref{ssec:StatsAndSenEst}, \Eref{e:h0SLW} presented the sensitivity
estimate of the sliding coherence window technique for a given (finite) amount of data, 
disregarding aspects of computational cost and essentially assuming unlimited 
computing power available.
The present section investigates the contrary case, finding
the amount of data which can be analyzed at limited (fixed) computational resources
and a given (fixed) coherence time baseline~$T$.

The computing cost $\zeta^{(1)}$ of a standard two-stage hierarchical search per
a certain volume of phase parameter space searched can always 
be written as a sum in terms of implementation-specific 
constants $\zeta_{\textrm{\tiny COH}}$ and $\zeta_{\textrm{\tiny INCOH}}$ 
pertaining to the coherent and incoherent combination stage,
respectively, as
\begin{equation}
  \zeta^{(1)} = \left(\zeta_{\textrm{\tiny COH}} + \zeta_{\textrm{\tiny INCOH}}\,\gamma^{(1)} \right)\, N^{(1)} \,,
  \label{e:zeta1}
\end{equation}
where $\gamma^{(1)}$~denotes the so-called refinement factor~\cite{pletsch:scmetric} 
of the incoherent combination stage,
and $N^{(1)}$ are the number of segments coherently analyzed.

The computational cost $ \zeta^{(q)}$ of the sliding coherence window technique
involves $q$ times more summations at the incoherent combination stage, thus,
\begin{equation}
  \zeta^{(q)} = \left( \zeta_{\textrm{\tiny COH}} + \zeta_{\textrm{\tiny INCOH}}\,\gamma^{(q)}\,q \right) \,N^{(q)} \,.
  \label{e:zetaq}
\end{equation}

The total amount of data which can be analyzed 
at the fixed computational expense using the standard hierarchical search scheme
is taken as \mbox{$T_{\textrm{\tiny data}}^{(1)} = T\,N^{(1)}$}.
In analogy, the amount of data that can be searched at given computing cost
using the sliding coherence window approach is 
\mbox{$T_{\textrm{\tiny data}}^{(q)} = T\,N^{(q)}$}, 
for the same coherent time baseline~$T$.

The sensitivity of the standard hierarchical search scheme
follows \mbox{$h_0^{(1)}\propto(T^{(1)}_{\textrm{\tiny data}}\; T)^{-1/4}$}.
Accordingly, the sensitivity of the 
sliding coherence window technique given in \Eref{e:h0SLW} scales as
\mbox{$h_0^{(q)}\propto (T^{(q)}_{\textrm{\tiny data}}\; T)^{-1/4}\; (2-1/q)^{-1/4}$}.
Thus, we define the sensitivity ratio by \mbox{$r \equiv h_0^{(1)} / h_0^{(q)}$},
which takes the form
\begin{equation}
   r = \left(\frac{T^{(q)}_{\textrm{\tiny data}}}{T^{(1)}_{\textrm{\tiny data}}}\right)^{1/4} \, 
    \left(2-\frac{1}{q}\right)^{1/4} \,.
   \label{e:ratior1}
\end{equation}
At equal total computing cost, \mbox{$\zeta^{(q)}=\zeta^{(1)}=\zeta$},
inverting \Esref{e:zeta1} and \eref{e:zetaq} for \mbox{$T^{(1)}_{\textrm{\tiny data}}$}
and \mbox{$T^{(q)}_{\textrm{\tiny data}}$}, respectively, one obtains
\begin{equation}
    r = \left(\frac{2}{q}-\frac{1}{q^2}\right)^{1/4} \, \left( \frac{\sqrt{1+4 q \theta}-1}{\sqrt{1+4 \theta}-1} \right)^{1/4} \,,
   \label{e:ratior2}
\end{equation}
where the constant $\theta$ has been defined as
\begin{equation}
   \theta \equiv \frac{\zeta \; \zeta_{\textrm{\tiny INCOH}}  }{\zeta_{\textrm{\tiny COH}}^2}\,,
   \label{e:theta}
\end{equation}
and the refinement factors have 
been approximated by \mbox{$\gamma^{(q)}\approx N^{(q)}$},
assuming the search includes at most one spindown 
parameter (cf. Ref.~\cite{pletsch:scmetric}).

\begin{figure}[t]
	\includegraphics[width=\columnwidth]{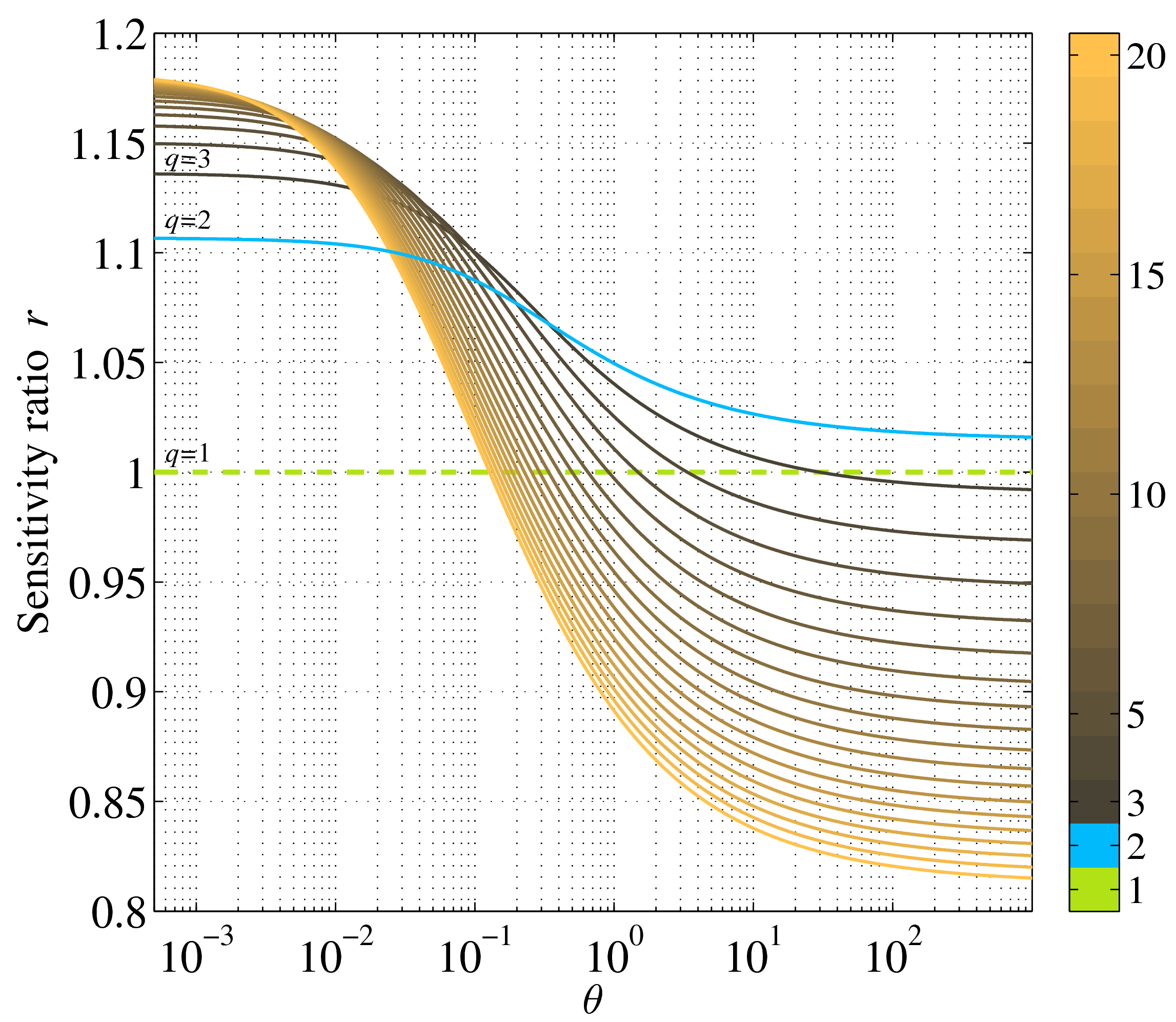}
	\caption{Estimated sensitivity ratio~$r$ as introduced in \Eref{e:ratior2} of the 
	standard hierarchical search and the sliding coherence 
	window technique at fixed total computational cost and given~$T$, 
	shown as a function of~$\theta$ defined in \Eref{e:theta}.
	The different curves correspond to different values 
	of $q$ (number of subsegments) as indicated
	by the shaded [color] bar. The dashed horizontal line corresponds 
	to \mbox{$q=1$}, for which the search methods coincide.
	\label{f:seniratio}}
\end{figure}

Figure~\ref{f:seniratio} illustrates the sensitivity ratio~$r$ of \Eref{e:ratior2} 
as a function of $\theta$ and for different values of~$q$. 
Two regimes are identified where $r$ is slowly changing:
when $\theta$ is either very small or very large. This can be understood as follows.
The specific value of $\theta$ depends on the code implementation, 
manifested in the two constants $\zeta_{\textrm{\tiny COH}}$ and $\zeta_{\textrm{\tiny INCOH}}$.
Thus, in the two extreme cases where one constant is much larger than the other, 
the two different limits of~$r$ result.

First, if the implementation is such that the coherent part dominates the computing cost
($\zeta_{\textrm{\tiny COH}} \gg \zeta_{\textrm{\tiny INCOH}} $), this
implies that $\theta$ is very small. Then the sensitivity ratio is described by
\begin{equation}
   \lim_{\theta \to 0} r = \left(2-\frac{1}{q}\right)^{1/4} =  \left(1+\coho\right)^{1/4}\,,
   \label{e:limr1}
\end{equation}
which is the same improvement factor as given by \Eref{e:h0SLWcoho}.

On the other hand, if the incoherent part  is the most computationally 
intensive ($\zeta_{\textrm{\tiny INCOH}} \gg \zeta_{\textrm{\tiny COH}} $), 
$\theta$ takes a very large value. 
In this case, the sensitivity ratio is described by
\begin{equation}
   \lim_{\theta \to \infty} r = \left(2-\frac{1}{q}\right)^{1/4} q^{-1/8}\,.
      \label{e:limr2}
\end{equation}

It is interesting to note that, only for \mbox{$q=2$}, the sensitivity ratio is \emph{always} 
greater than $1$, implying that in this case the sliding coherence window technique 
should always be more sensitive than the standard scheme at given~$T$ 
and fixed computational cost. 

In the current \mbox{Einstein@Home}~\cite{eahurl} analysis, 
$\theta$ is approximately of order $10^{-2}$. 
Thus,  the situation in this case is rather comparable to the regime described by \Eref{e:limr1}. 
Hence, the \mbox{Einstein@Home} search sensitivity will certainly benefit 
from employing the sliding coherence window technique.

\section{Conclusion\label{sec:Conclusion}}

In summary, a novel hierarchical strategy to search
for prior unknown continuous gravitational-wave sources
has been presented, exploiting a sliding coherence window.
The standard hierarchical search scheme divides the data 
into~$N$ nonoverlapping segments that are coherently analyzed,
and subsequently matched-filter outputs are combined incoherently.
Thereby, the duration of one segment defines
the maximum time span of coherence.
In contrast, the presented sliding coherence window approach 
divides each of the $N$~data segments into $q$ subsegments,
which are thus \emph{shorter} than the desired maximum 
coherence length~$T$ (size of the coherence window). 
This permits the efficient combination of matched-filter outputs
from all subsegments in a ``sliding-window''
fashion: If subsegments are closer than~$T$,
they are combined coherently; otherwise, they are combined incoherently.

As a result, the estimated search sensitivity of the sliding coherence window approach 
is considerably superior compared to the standard hierarchical scheme,
while using the \emph{same} number of  coarse- and fine-grid templates to cover 
the search parameter space.
At a given value of $T$, the sensitivity improvement in terms of minimum 
detectable gravitational-wave amplitude~$h_0$  scales with the fourth root 
of \mbox{$N(2-1/q)$} for a contiguous data set.
Since for the standard hierarchical method~$q=1$, 
to achieve the same sensitivity as the sliding coherence window technique
between \mbox{$50-100\,\%$} more data (to increase~$N$ accordingly) 
would have to be analyzed.
Realistic Monte-Carlo simulations have been carried out
confirming the sensitivity enhancement.

The sensitivity improvement can also be expressed in terms of
the average coherence overlap~$\coho$ between successive sliding steps.
In particular, if the data set has gaps in time, $\coho$ can be a 
useful figure of merit.  The estimated sensitivity improvement of the sliding coherence 
window technique over the standard scheme  scales as the fourth root of \mbox{$(1+\coho)$}.

In addition, the sensitivity has also been compared at fixed computational cost.
When the computing cost of the coherent stage dominates,
the above sensitivity improvement holds.
In the case where the computational cost of the incoherent stage
dominates, the sensitivity improvement can fade away, depending on the search setup.
However, it is estimated that, when the chosen 
coherence window is equal to the length of~$2$~subsegments, the sensitivity is always
superior to the standard method at about the same computational expense.
In general, the search setup (including the choice of~$q$) will have to be 
sensitivity optimized at the given computational resources 
and the software at hand, as well as for the data available. 


However, further topics are planned to be investigated in future work.
One of these aspects concerns the efficient implementation of the proposed technique
while exploiting the FFT algorithm. Moreover, an optimal weighting scheme between 
the coherent matched-filter outputs from different subsegments could be further studied,
taking into account correlations between these.

The sliding coherence window approach is envisioned to be employed 
by the Einstein@Home~\cite{eahurl} project to further improve the search 
sensitivity of all-sky surveys for unknown isolated CW sources~\cite{pletsch:GCT}. 
The proposed approach should also be extensible to CW searches for sources in 
binary systems~\cite{Messenger2010}.
With suitable modification, the method might also have applicability in further related areas, 
for instance, regarding computationally limited searches for prior unknown
radio~\cite{Ransom2002Fourier}, X-ray~\cite{X-rayBinarySearches1994}, and 
gamma-ray pulsars~\cite{Chandler2001,TimeDiffTech2006,FermiLAT16pulsars2009,FermiLAT8pulsars2010}.

\section{Acknowledgments}
I am grateful to \mbox{Bruce Allen}, \mbox{Badri Krishnan},  \mbox{Reinhard} Prix,
and  \mbox{Karl} Wette for numerous valuable discussions. 
The support of the Max-Planck-Society is gratefully acknowledged.
This document has been assigned LIGO Document Number \mbox{\dcc}.


\bibliography{MLESLCW}

\end{document}